\documentclass[pre,aps,preprint]{revtex4}
\usepackage{graphicx}
\usepackage{amsfonts}
\usepackage{amssymb}
\usepackage{amsmath}
\usepackage{textcomp}

\usepackage{color}
\usepackage{psfrag}

\usepackage{epsfig}

\begin{document}

\title{A Packing of Truncated Tetrahedra that Nearly Fills All of Space}


\author{Yang Jiao,$^{1}$ and Salvatore Torquato$^{1,2}$}
\email{torquato@princeton.edu}
\affiliation{$^{1}$Princeton Institute of the Science and Technology of Materials,
$^{2}$Department of Chemistry and Physics, Princeton Center for Theoretical Science,
Princeton University, Princeton, New Jersey 08544, USA}

\begin{abstract}

Dense polyhedron packings are useful models of a variety of condensed
matter and biological systems and have intrigued scientists mathematicians for centuries.
Recently, organizing principles for the types of structures
associated with the densest polyhedron packings have been put forth. 
However, finding the maximally dense packings for specific shapes
remains nontrivial and is a subject of intensive research. Here, 
we analytically construct the densest known packing
of truncated tetrahedra with $\phi=207/208=0.995~192\ldots$,
which is amazingly close to unity and strongly implies that
this packing is maximally dense. This construction is based on a generalized organizing
principle for polyhedra that lack central symmetry.
Moreover, we find that the holes in this putative optimal packing
are small regular tetrahedra, leading to a new tiling of space
by regular tetrahedra and truncated tetrahedra. We also numerically study the equilibrium melting
properties of what apparently is the densest packing of truncated
tetrahedra as the system undergoes decompression. Our simulations
reveal two different stable crystal phases, one at high densities and
the other at intermediate densities, as well as a first-order liquid-crystal
phase transition.

\end{abstract}

\maketitle




In general, a packing is defined as a large collection of nonoverlapping solid objects
(particles) in $d$-dimensional Euclidean space $\mathbb{R}^d$.
Associated with a packing is the density $\phi$ defined as the fraction of
space $\mathbb{R}^d$ covered by the particles. Dense packings are useful models of a variety of condensed
matter systems, including liquids, glasses and crystals \cite{Ber65,Za83,Ch00,To02}, granular
media \cite{Ed94,To02}, and heterogeneous materials \cite{To02}.
Understanding how nonspherical particles pack is a first
step toward a better understanding of how biological cells pack \cite{Li01,Pu03,Ge08}.
Probing the symmetries and other mathematical properties of
the densest packings is a problem of interest in discrete geometry and number theory \cite{ConwayBook, Henry}.

The Platonic and Archimedean polyhedra possess beautiful symmetries and arise
in many natural and synthetic structures. An original investigation of dense
packings of regular tetrahedra, one of the five Platonic solids, by Conway and Torquato \cite{Co06}
has spurred a flurry of subsequent research activity, including studies to obtain
dense packing of hard polyhedra \cite{ToJi09, ToJi09b, Haji09, Ka10, ToJi10, Ch10}
and to understand the phase behavior of colloidal systems made of such solid objects \cite{Haji09, Aga11}.
General organizing principles have been established for densest packings of polyhedral shapes \cite{ToJi09, ToJi09b}.
For centrally symmetric Platonic and Archimedean solids, it has been conjectured that the densest packings can be
achieved by arranging the shapes on an appropriate Bravais lattice (see definition below)
with the same orientation \cite{ToJi09}.(A centrally symmetric solid has
a center of inversion symmetry.) For non-centrally symmetric shapes, the
optimal packings are generally not Bravais-lattice packings \cite{ToJi09b}. For example,
a tetrahedron lacks central symmetry, and the densest known packings
of such a polyhedron possess a fundamental cell (defined below)
with four tetrahedra that form a centrally symmetric repeat unit \cite{Ka10, ToJi10, Ch10}.

The Archimedean analog of the regular tetrahedron is the truncated tetrahedron,
whose densest packing has been shown to be a non-Bravais lattice packing \cite{Co06, Gr11}.
An Archimedean truncated tetrahedron has four hexagonal faces and four triangular faces, as obtained by truncating
the corners (small tetrahedra) of a regular tetrahedron (see Fig. 1(a)).
In particular, a truncated tetrahedron does not possess central symmetry
and its dihedral angle (i.e., the angle between two faces) is the same as that of tetrahedron
 $\theta = \cos^{-1}(1/3) \approx 0.392 \pi$. Since $\theta$ is not a sub-multiple
of $2 \pi$, one cannot tile the three-dimensional Euclidean space $\mathbb{R}^3$ with truncated tetrahedra.
Betke and Henk showed that the optimal Bravias lattice packing of
truncated tetrahedra possesses a density $\phi = 207/304 = 0.680~921\ldots$ \cite{Be00}.
Conway and Torquato constructed a dense packing of truncated tetrahedra with density
$\phi=23/24 = 0.958~333\ldots$  \cite{Co06}, which proved that the optimal packing
must be a non-Bravais lattice packing. Recently, de Graaf, van Roij and Dijkstra \cite{Gr11}
showed via numerical simulations that truncated tetrahedra can pack at least as dense as $0.988$,
suggesting the existence of even denser packings. Indeed, we find an exact construction
for the densest known packing of truncated tetrahedra with density $\phi= 207/208=0.995~192\ldots$
which nearly fills all of space. This construction is based on a generalized organizing
principle for polyhedra that lack central symmetry.

Some important definitions are in order here before describing our new packing.
A lattice $\Lambda$ in $\mathbb{R}^3$ is an infinite set of points
generated by a set of discrete translation operations defined by integer
linear combinations of a basis of $\mathbb{R}^3$, i.e., ${\bf a}_1, {\bf a}_2$ and ${\bf a}_3$ \cite{Ch00}.
The vectors ${\bf a}_i$ ($i=1,2,3$) are called lattice vectors for $\Lambda$.
A (Bravais) lattice packing associated with $\Lambda$ is one in which the centroids (geometric centers)
of the nonoverlapping particles are located at the points of $\Lambda$, each oriented in
the same direction. The space $\mathbb{R}^3$
can then be geometrically divided into identical regions $F$ called
fundamental cells, each of which contains just the centroid
of one particle. Thus, the density of a lattice packing is given by
\begin{equation}
\label{eq01}
\phi= \frac{V_p}{\mbox{Vol}(F)},
\end{equation}
where $V_p$ is the volume of a  particle and $\mbox{Vol}(F) = |{\bf a}_1 \times {\bf a}_2 \cdot {\bf a}_3|$
is the volume of a fundamental cell. A periodic (non-Bravais lattice) packing is
obtained by placing a fixed nonoverlapping configuration of $N$
particles (where $N\ge 1$) with arbitrary orientations in
each fundamental cell of a lattice $\Lambda$. Thus, the packing is
still periodic under translations by $\Lambda$, but the $N$
particles can occur anywhere in the chosen cell subject to the nonoverlap condition. The density of a
periodic packing is given by
\begin{equation}
\label{eq02}
\phi=\frac{N V_p}{\mbox{Vol}(F)}.
\end{equation}

Since a truncated tetrahedron lacks central symmetry, its optimal packing can
only be a non-Bravais lattice packing. Based on our principles developed for
determining the densest polyhedron packings \cite{ToJi09, ToJi09b, ToJi10},
we argue more generally that the fundamental cell of the optimal packing of truncated tetrahedra
should contain a simple compound object composed of truncated tetrahedra that itself
is centrally symmetric. The fact that the aforementioned Conway-Torquato
packing (henceforth referred to as ``CT packing'') possesses such a fundamental cell
containing a centrally symmetry dimer (defined below) of truncated tetrahedra
with high density $\phi = 23/24 = 0.958~333\ldots$ \cite{Co06} suggests
that this packing can be used as a starting point to find the optimal packing.
Indeed, in the ensuing discussion, we provide a construction of the densest known packing of
truncated tetrahedra by optimizing the CT packing.

It is convenient to describe a truncated tetrahedron from
its associated tetrahedron with vertices labeled A, B, C and D [see Fig.~1(a)].
The centers ${\bf p}_1$, ${\bf p}_2$ and ${\bf p}_3$ of the hexagonal
faces of the truncated tetrahedron can be expressed as
\begin{equation}
\label{eq03}
\begin{array}{c}
{\bf p}_1 = \frac{1}{3}({\bf v}_{\mbox{\tiny A}}+{\bf v}_{\mbox{\tiny B}}+{\bf v}_{\mbox{\tiny D}})-{\bf O},\\
{\bf p}_2 = \frac{1}{3}({\bf v}_{\mbox{\tiny A}}+{\bf v}_{\mbox{\tiny C}}+{\bf v}_{\mbox{\tiny D}})-{\bf O}, \\
{\bf p}_3 = \frac{1}{3}({\bf v}_{\mbox{\tiny B}}+{\bf v}_{\mbox{\tiny C}}+{\bf v}_{\mbox{\tiny D}})-{\bf O},
\end{array}
\end{equation}
where ${\bf v}_{i}$ ($i=\mbox{A, B, C, D}$) are the vectors associated with
the vertices and the origin ${\bf O}=\frac{1}{3}({\bf v}_{\mbox{\tiny A}}+{\bf v}_{\mbox{\tiny B}}+{\bf v}_{\mbox{\tiny C}})$.
In the aforementioned CT packing \cite{Co06}, each fundamental cell
contains two truncated tetrahedra, making a perfect contact through one of the hexagonal
faces of each particle, which are center-inversion images of each other through ${\bf O}$ [see Fig.~1(b)].
Such a repeat unit is centrally symmetric and can be considered to be a regular rhombohedron with two sharper
corners (with in-face angle $\pi/3$) truncated. The truncated rhombohedron
has six pentagonal faces and two triangular faces [Fig.~1(b)]. The CT packing then
corresponds to removing the sharp corners (tetrahedra with half edge-length of
the rhombohedron) of each rhombohedron in its tiling [see Fig.~1(c)],
leading to $\phi=23/24$ with the lattice vectors:
\begin{equation}
\label{eq1}
{\bf a}_1 = -2{\bf p}_3, \quad {\bf a}_2 = -2{\bf p}_2, \quad {\bf a}_3 = 2{\bf p}_1.
\end{equation}
Both the truncated-tetrahedron dimer (i.e., the truncated regular rhombohedron)
and the CT packing possess 3-fold rotational symmetry, with the symmetry
axes being the long body-diagonal of the rhombudedron. Each dimer makes
contacts with six neighbors through its six pentagonal faces.

\begin{figure}
\begin{center}
$\begin{array}{c}
\includegraphics[height=6.75cm,keepaspectratio]{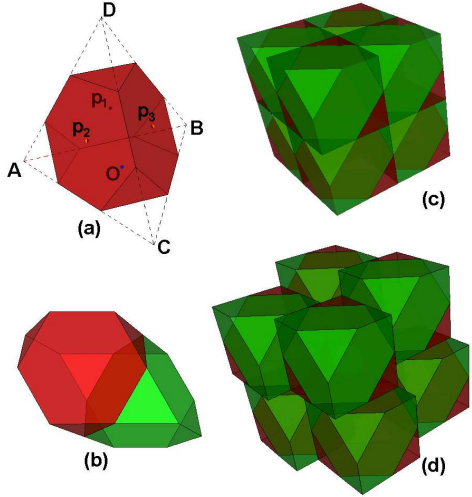} \\
\end{array}$
\end{center}
\caption{Truncated tetrahedron and portions of its packings. (a) A truncated
tetrahedron with its associated tetrahedron shown. The labels are described in the text.
(b) A dimer (pair) of truncated tetrahedra making perfect contact through
a hexagonal faces. Such a dimer is also a truncated regular
rhombohedron. (c) A portion of the Conway-Torquato packing of
truncated tetrahedra with a density $\phi=23/24 = 0.958~333\ldots$. The holes
in this packing are regular tetrahedra with the same edge length as the truncated
tetrahedra. (d) A portion of the densest known packing of truncated tetrahedra with
$\phi=207/208=0.995~192\ldots$. The holes in this densest known packing are
regular tetrahedra whose edge length is 1/3 of that of truncated tetrahedra.}
\label{fig_1}
\end{figure}

It is noteworthy that the CT packing is not
``collectively'' jammed and can be continuously deformed
until the density reaches a (local) maximum. Following Torquato and Stillinger
\cite{To01}, a packing is {\it locally} jammed if no particle in
the system can be translated while fixing the positions of all
other particles. A {\it collectively} jammed packing is a locally
jammed packing such that no subset of particles can simultaneously
be continuously displaced so that its members move out of contact
with one another and with the remainder set. A packing is {\it
strictly} jammed if it is collectively jammed and all globally
uniform volume non-increasing deformations of the system boundary
are disallowed by the impenetrability constraints. Readers are
referred to Ref.\cite{To01} for further details.

\begin{figure}
\begin{center}
$\begin{array}{c}
\includegraphics[height=2.95cm,keepaspectratio]{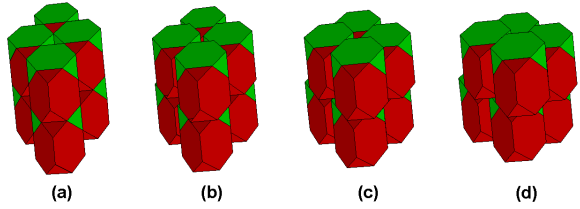} \\
\end{array}$
\end{center}
\caption{Structures along the continuous deformation path that 
brings the Conway-Torquato packing of truncated tetrahedra with $\phi =
23/24$ to the putative optimal packing with $\phi = 207/208$. The
associated deformation parameter $\gamma$ for each packing is (a)
$\gamma = 0$ (the Conway-Torquato packing),  (b) $\gamma = 2/27$,  (c) $\gamma = 4/27$ and (d)
$\gamma = 2/9$ (our current putative optimal packing).} \label{fig_2}
\end{figure}

The truncated-tetrahedron dimers in the CT packing can slide
relative to one another along directions of certain edges,
while the 3-fold rotational symmetry of the packing is maintained (see Fig. 2).
Such a deformation can be achieved via continuously varying the lattice vectors:
\begin{equation}
\label{eq2}
\begin{array}{c}
{\bf a}_1 = -2{\bf p}_3+\gamma({\bf v}_{\mbox{\tiny D}}-{\bf v}_{\mbox{\tiny C}}), \\
{\bf a}_2 = -2{\bf p}_2+\gamma({\bf v}_{\mbox{\tiny D}}-{\bf v}_{\mbox{\tiny A}}),  \\
{\bf a}_3 = 2{\bf p}_1+\gamma({\bf v}_{\mbox{\tiny B}}-{\bf v}_{\mbox{\tiny D}}),
\end{array}
\end{equation}
where $\gamma \in [0, 2/9]$ is the deformation parameter determined by the
nonoverlapping constraints. For $\gamma=0$,
one has the CT packing; and $\gamma = 2/9$ corresponds to the densest known
packing of truncated tetrahedra. By explicitly choosing a
set of coordinates for ${\bf v}_{i}$ ($i=\mbox{A, B, C, D}$), e.g.,
${\bf v}_{\mbox{\tiny A}} = (-1, ~-1, ~-1)$, ${\bf v}_{\mbox{\tiny B}} = (1, ~1, ~-1)$,
${\bf v}_{\mbox{\tiny c}} = (-1, ~1, ~1)$, ${\bf v}_{\mbox{\tiny D}} = (1, ~-1, ~1)$,
the packing density can be obtained using Eqs.~\ref{eq02}, \ref{eq03} and \ref{eq2}:
\begin{equation}
\phi=\frac{2 V_p}{\mbox{Vol}(F)} = \frac{2\times(184/81)}{76544/16767} = \frac{207}{208} = 0.995~192\ldots.
\end{equation}

The aforementioned continuous deformation which leads to the putative optimal
packing from a sub-optimal one can be better understood by appealing
to a two-dimensional analog, as illustrated in Fig. 3.
In particular, consider a rhombus with interior angles $\pi/3$ and $2\pi/3$
whose optimal packing completely fills the plane without any gaps, i.e., with $\phi = 1$. This is usually called
a \textit{tiling} or \textit{tessellation} of the plane. Now imagine that the two sharp corners (i.e., with angles $\pi/3$)
of the rhombus are cut so that the remaining shape is a regular hexagon.
It is obvious that the resulting packing of hexagons is not optimal (with $\phi = 3/4$). One can
collectively deform this packing (as shown in Fig.~3(c)) to achieve a new packing
of hexagons with $\phi = 1$.

\begin{figure}
\begin{center}
$\begin{array}{c}
\includegraphics[height=2.05cm,keepaspectratio]{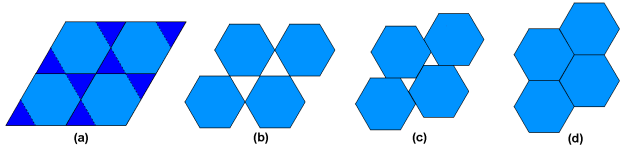} \\
\end{array}$
\end{center}
\caption{Structures along the continuous deformation path that brings 
a low-density packing of truncated rhombi (i.e., regular hexagons ) with $\phi = 3/4$ to 
the optimal triangular-lattice packing with $\phi = 1$ in two dimensions.
(a) The optimal packing of rhombi which completely fills the plane, i.e., $\phi=1$.
(b) The truncated rhombi (i.e., regular hexagons) arranged on the optimal
lattice for rhombi with $\phi = 3/4$. (c) Deformation of the packing of hexagons which
leads to denser packings. (d) The optimal packing of hexagons with $\phi = 1$ obtained by
the continuous deformation.}
\label{fig_3}
\end{figure}

In the densest known packing, since the dimers have slided relative to each other,
each dimer now contacts two neighbors through one pentagonal face.
Moreover, each dimer makes contacts with two additional neighbors along the
direction of its long body-diagonal (i.e., ${\bf a}_1+{\bf a}_2-{\bf a}_3$)
through the two triangular faces [see Fig.~1(d)]. Thus, each dimer has 14 face-to-face contacts
in the densest known packing. We note that the formation of the additional
triangular-face contacts prevents further deformations of the packing.
The contacting equilateral triangular faces are center-inversion
images of each other and share a common face center. This leads
to six tetrahedron holes (with edge length 1/3 of that of truncated tetrahedra)
associated with each triangular face-face contact and thus, each
truncated-tetrahedron dimer. Let the volume of the holes $V_H=1$, then
the volume of a truncated tetrahedron $V_T=621$. Alternatively, the density can be easily obtained via
\begin{equation}
\label{eq3}
\phi=\frac{V_T+V_T}{V_T+V_T+6V_H}=\frac{621+621}{621+621+6}=\frac{207}{208}
\end{equation}
which is amazingly close to unity, given the fact that truncated tetrahedra can not tile $\mathbb{R}^3$.
Numerical maximization methods have been employed to verify that the densest known
packing is indeed optimal among packings with similar structures. In fact,
its remarkably high density, highly symmetric structure and the
numerical maximization results all suggest that this packing would be
the optimal among all packings of truncated tetrahedra.

We note that by inserting regular tetrahedra of proper size
into the holes of the optimal packing of the truncated tetrahedra, one
completely fills the space without any gaps. This leads to a new tiling (or tessellation)
of $\mathbb{R}^3$ by regular tetrahedra and truncated tetrahedra.
Since the holes in the CT packing are also regular
tetrahedra, a tiling associated with this packing can be obtained in a similar way.
Interestingly, it has recently been shown that the optimal packing of
regular octahedra, which can be obtained by continuous deforming a sub-optimal
face-centered-cubic packing of octahedra, corresponds to a new tiling of $\mathbb{R}^3$ by regular
tetrahedra and octahedra \cite{Con11}. In fact, each octahedron packing
in the family of packings obtained by a continuous deformation is associated
with a tiling of tetrahedra and octahedra, unlike the situation for truncated tetrahedra where
only the two ``extremes'' of the deformation correspond to tilings.

Once again, we see the important role that central symmetry plays in dense packings:
though truncated tetrahedra are not centrally symmetric, they form centrally symmetric
dimers, which then densely pack on a Bravais lattice. This is also the case
for the densest tetrahedron packing, whose centrally symmetric unit has four particles
forming two dimers. This suggests a generalization of the organizing principle we
proposed for centrally symmetric Platonic and Archimedean solids. Specifically,
the densest packings of convex polyhedra with equivalent principal axes are either a Bravais lattice packing
of the polyhedra themselves that possess central symmetry or
a Bravais lattice packing of centrally symmetric compound solids that are made of
the polyhedra that lack central symmetry.

\begin{figure}
\begin{center}
$\begin{array}{c}
\includegraphics[height=5.75cm,keepaspectratio]{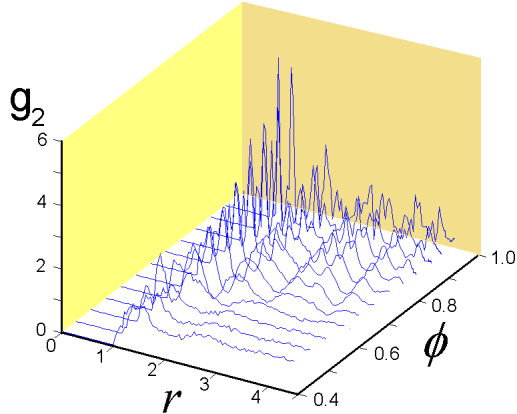} \\
\end{array}$
\end{center}
\caption{Pair correlation function $g_2(r)$ associated with the centroids
of the truncated tetrahedra at different densities in the range from 0.99 to 0.38 
during the melting of the highest-density crystal. At very high densities, 
$g_2(r)$ is a long-ranged function.  
Below $\phi \approx 0.53$, $g_2(r)$ suddenly changes from a long-ranged function to an exponential
decaying function, indicating the occurrence of a first-order crystal-liquid transition.}
\label{fig_4}
\end{figure}

Although a comprehensive study of the equilibrium phase behavior is beyond the
scope of the present paper, we conclude our investigation by examining
the equilibrium melting properties of our putative optimal packing of truncated tetrahedra.
At infinite pressure, the optimal packing of truncated tetrahedra constructed
here is the thermodynamic equilibrium phase for these particles.
Colloidal particles with shapes similar to that of truncated tetrahedra
have been fabricated via different techniques \cite{Wang97, Shao07, Seo08}.
Thus, it is useful to see to what extent this solid phase is stable under
finite pressures. We have carried out Monte-Carlo (MC) simulations \cite{Frenkel_Book}
to ``melt'' the optimal packing structure (i.e., the highest-density ``crystal'') via a decompression process.
In particular, a periodic simulation box containing $N=686$ particles is
employed, whose size and shape are allowed to change \cite{ToJi09, ToJi10}.
The volume of the simulation box is slowly increased to decrease
the pressure and density of the system. At each density, a total number of
$10~000~000$ MC trial moves are applied to each particle and $100~000$
trial volume-preserving deformations are applied to the simulation box to
equilibrate the systems. Equilibrium structural characteristics, such as the pair-correlation
function $g_2$ (see Fig. 4) and the number of dimers $n_2$, are collected.
We use such descriptors to gauge the remaining crystalline
order in the system during the decompression process.
We find that above $\phi \approx 0.68$, the crystal configurations associated with the optimal packing
of truncated tetrahedra is the stable solid phase for these particles.
When $0.53 < \phi < 0.68$, the crystal configurations associated with the
CT packing becomes the stable phase.
Below $\phi \approx 0.53$, the correlation function $g_2$ suddenly
changes from a long-ranged function to an exponential decaying function (see Fig. 4)
and $n_2$ quickly drops from $N/2$ to almost zero,
indicating the occurrence of a first-order crystal-liquid transition.


Our simulations suggest that the crystal phases associated with the optimal
packing and CT packing of truncated tetrahedra
are stable over a wide range of densities $\phi \in (0.53, 0.995)$ upon
melting (decompression). This wide range of stability for the crystal phase is due to
the fact that the dimers of truncated tetrahedra (in both the CT packing and
the putative optimal packing) fill space very efficiently.
This means that the free volume associated with crystals
of the truncated tetrahedra is readily maximized in the dimer arrangement,
leading to a lower free energy of the system.
Since a dimer is formed by a pair of truncated tetrahedra contacting through
the a common large hexagonal face, it is relative easy for
such local clusters to form in a relatively dense liquid. Once such
dimers nucleate, the system is expected to crystallize easily upon further compression.
Thus, we expect the phase diagram of truncated tetrahedra to involve
a single first-order liquid-solid phase
transition. Of course, the exact coexistence range of $\phi$ and
whether there are higher-order solid-solid phase transitions can be
precisely explored by carrying out free-energy calculations, which we intend to do in future work.

In summary, we have discovered a packing of truncated
tetrahedra that nearly fills all of space, i.e., $\phi= 207/208 = 0.995~192\ldots$,
via exact analytical construction. We are not aware of any packing of a nontiling regular or
semi-regular polyhedron with packing density that is nearly unity.
While a rigorous proof that this packing is indeed optimal (maximally dense)
is highly nontrivial, the fact that its packing density is close
to unity in conjunction with our generalized organizing principle
leads us to conclude that it is likely optimal. This also
leads us to a new organizing principle for polyhedral shapes that
lack central symmetry, which is the analog of the principle for centrally
symmetric shapes that we proposed in Ref.~[11].


\begin{acknowledgments}
 Y. J. thanks his parents for helping make the models of truncated tetrahedra.
This work was supported by the National Science Foundation under Award Numbers
DMS-0804431 and DMR-0820341.
\end{acknowledgments}

\end{document}